\RequirePackage{fix-cm}
\documentclass[smallextended]{svjour3}       
\smartqed  
\usepackage{graphicx}
 \usepackage{mathptmx}      
\usepackage[usenames,dvipsnames]{color}
\usepackage{latexsym,a4wide}
\usepackage[hang,nooneline]{subfigure}
\newcounter{saveeqn}
\newcommand{\alpheqn}
{ \setcounter{saveeqn}{\value{equation}}\stepcounter{saveeqn}
 \setcounter{equation}{0}
 \renewcommand{\theequation}{\mbox{\arabic{saveeqn}\alph{equation}}}}
\newcommand{\reseteqn}
{ \setcounter{equation}{\value{saveeqn}}
 \renewcommand{\theequation}{\arabic{equation}}}
 \journalname{General Relativity and Gravitation}
\begin{document}

\title{Boson Stars in a Theory of Complex Scalar Field coupled to Gravity}


\author{Sanjeev Kumar \and Usha Kulshreshtha \and \\Daya Shankar Kulshreshtha}


\institute{Sanjeev Kumar \at
              Department of Physics and Astrophysics,  
			  University of Delhi, Delhi-110007, India\\
			  \email{sanjeev.kumar.ka@gmail.com}           
           \and
           Usha Kulshreshtha \at
              Department of Physics,
              Kirorimal college,
              University of Delhi, Delhi-110007, India\\
              \email{ushakulsh@gmail.com}
              \and
              Daya Shankar Kulshreshtha\at
              Department of Physics and Astrophysics, 
			  University of Delhi, Delhi-110007, India\\
			  \email{dskulsh@gmail.com}           
}

\date{Received: date / Accepted: date}

\maketitle

\begin{abstract}
We study boson stars in a theory of complex scalar field coupled to Einstein gravity with the potential: $V(|\Phi|) := m^{2} |\Phi|^2 +2 \lambda |\Phi|$ (where $m^2$ and $\lambda$ are positive constant parameters). This could be considered either as a theory of massive complex scalar field coupled to gravity in a conical potential or as a theory in the presence of a potential which is an overlap of a parabolic and a conical potential. We study our theory with positive as well as negative values of the cosmological constant $\Lambda$. Boson stars are found to come in two types, having either ball-like or shell-like charge density.  We have studied the properties of these solutions and have also determined their domains of existence for some specific values of the parameters of the theory. Similar solutions have also been obtained by Hartmann, Kleihaus, Kunz, and Schaffer, in a V-shaped scalar potential. 

\end{abstract}
\vspace{0.5cm}
A study of boson shells and boson stars in scalar electrodynamics with a self-interacting complex scalar field $\Phi$ coupled to Einstein gravity is of a very wide interest in the gravity theories\cite{sanjeev:2014cqg}-\cite{Brihaye:2013hx} . Hartmann, Kleihaus, Kunz, and Schaffer (HKKS) \cite{Hartmann:2012da,Hartmann:2013kna} have recently studied boson stars in a theory of complex scalar field coupled to Einstein gravity  in a V-shaped scalar potential: $V(\Phi \Phi^* )\equiv V(|\Phi|) = \lambda_c |\Phi|$ (where $\lambda_{c}$ is a constant). They have found that the boson stars come in two types, having either ball-like or shell-like charge density. They have studied the properties of these solutions and have also determined their domains of existence.

Actually, the boson stars represent localized self-gravitating solutions \cite{Feinblum:1968,Kaup:1968zz,Ruffini:1969qy}, that have been considered in many different contexts \cite{Jetzer:1991jr,Lee:1991ax,Mielke:1997re,Mielke:2000mh,Liebling:2012fv}. To obtain boson stars, typically a complex scalar field $\Phi$ is considered.
The U(1) invariance of the theory then provides a conserved Noether current.

The properties of the boson stars depend strongly on the self-interaction employed. In particular, as discussed by Lee and collaborators \cite{Friedberg:1976me} and by Coleman \cite{Coleman:1985ki}, the existence of a flat space-time limit of these localized solutions, i.e., the existence of $Q$-balls, puts constraints on the types of self-interaction possible.

Arodz and Liz showed, that besides $Q$-balls another type of localized solution could also arise in flat space, when a V-shaped self-interaction is employed in the presence of a gauge field \cite{Arodz:2008jk,Arodz:2008nm,Arodz:2012zh}. In this type of solution, the energy density is no longer ball-like as in the case of $Q$-balls, but it is instead shell-like. Here the scalar field is finite in a compact region of space, $r_i \le r \le r_o$, where $r_i$ and $r_o$ are the inner and outer radii of the shell, respectively. Therefore these solutions have been termed $Q$-shells. $Q$-shells, in fact, arise smoothly from $Q$-balls, when the parameters are varied appropriately. 

Inspired by these findings, Kleihaus, Kunz, L\"ammerzahl and List \cite{Kleihaus:2009kr,Kleihaus:2010ep} showed, that these shell-like solutions exist even when one introduces explicitly the gravitational field  into the theory (represented by the Ricci scalar $R$). These authors have explored the domains of existence of compact boson stars and boson shells in the model of scalar electrodynamics with a V-shaped self-interaction potential, coupled to gravity. Subsequently they have also considered the boson shells, which do not have an empty inner region $r<r_i$, but instead they harbour a Schwarzschild black hole or a Reissner-Nordstr\"om black hole in the region $r<r_i$. For the details of related studies we refer to the work of Refs. \cite{Hartmann:2012da,Hartmann:2013kna,Kleihaus:2009kr,Kleihaus:2010ep}.

Another topical development in this field is the study of such solutions in the presence of a cosmological constant. This is amply supported by the studies on AdS/CFT correspondence  \cite{Maldacena:1997re,Witten:1998qj}. Not only that, Astefanesei and Radu \cite{Astefanesei:2003qy}, and others \cite{Prikas:2004yw,Hartmann:2012wa,Radu:2012yx,Hartmann:2012gw,Brihaye:2013hx}, have shown that such studies also remain meaningful even in the presence of a negative cosmological constant.

In the present work, we study boson stars in a theory of complex scalar field coupled to Einstein gravity with the potential $V(|\Phi|)$ defined by: 
\begin{equation}
V(|\Phi|) := m^{2} |\Phi|^2 + 2 \lambda |\Phi| \label{potential}
\end{equation}
where $m^2$ and $\lambda$ are positive constant parameters. 
This could be considered either as a theory of massive complex scalar field coupled to gravity in a conical potential or as a theory in the presence of a potential which is an overlap of a parabolic and a conical potential.

The action of the theory under consideration reads:
\begin{equation}
S=\int \left[ \frac{(R - 2\Lambda) }{16\pi G}+\mathcal L_M \right] \sqrt{-g} ~~ d^4 x  ~~;~~~~~    \mathcal L_M =- \left( \nabla_\mu \Phi \right)^* \left( \nabla^\mu \Phi \right)- V( \Phi \Phi^* )    \label{action}
\end{equation}
Here $\mathcal L_M$ is the matter Langrangian density, $R$ is the Ricci curvature scalar, $\Lambda$ is the Cosmological constant, $G$ is Newton's Gravitational constant, $g = det(g_{\mu\nu})$ where the metric tensor $g_{\mu\nu}$ is defined later, $\nabla_\mu$ denotes the covariant derivative and the asterisk in the above equations denote complex conjugation (here, $\mu,\nu$ = t, r, $\theta$, $\varphi$).

Also, to construct static spherically symmetric solutions we adopt the spherically symmetric metric with Schwarzschild-like coordinates
\begin{eqnarray}
 ds^2&=& g_{\mu\nu} dx^\mu dx^\nu = \biggl[ -A^2(r) N(r) dt^2 + N^{-1}(r) dr^2 +r^2(d\theta^2 + \sin^2 \theta \;d\varphi^2) \biggr]
\end{eqnarray}
Here the signature of the metric $g_{\mu\nu}$ is (-,+,+,+). The equations of Motion for the fields are obtained by variation of the action with respect to the metric and the matter fields
\begin{eqnarray}
 G_{\mu\nu}&\equiv& R_{\mu\nu}-\frac{1}{2}g_{\mu\nu}R = 8\pi G T_{\mu\nu} -\Lambda\,g_{\mu\nu}
\nonumber \\ 
\nabla_\mu\left(\sqrt{-g}  \nabla^\mu \Phi \right) &=&
    \sqrt{-g}\,m^2\Phi +\sqrt{-g}\,\lambda \frac{\Phi}{|\Phi|}
  \label{eq:phi}
 \end{eqnarray}
The equation of motion for the field $\Phi^{*}$ is obtained by the complex conjugation of the last equation. The stress-energy tensor $T_{\mu\nu}$ is given by ,
\begin{eqnarray}
T_{\mu\nu} 
&=& g_{\mu\nu}{\mathcal{L}}_M -2 \frac{\partial \mathcal{{L}}_M}{\partial g^{\mu\nu}}\nonumber\\
&=& \biggl[  (\nabla_\mu \Phi)^* (\nabla_\nu \Phi) + (\nabla_\nu \Phi)^* (\nabla_\mu \Phi)- g_{\mu\nu}  (\nabla_\alpha \Phi)^* (\nabla_\beta \Phi)    g^{\alpha\beta}-  g_{\mu\nu}\; V( |\Phi|) \biggr] ~~~
  \label{eq:tmunu}
\end{eqnarray}

We now obtain
\begin{eqnarray}
G_t^t &=& \biggl[ -\frac{1}{r^2}\left[r\left(1-N\right)\right]' \biggr] ~~,~~
G_r^r = \biggl[ \frac{2 r A' N -A\left[r\left(1-N\right)\right]'}{A\ r^2} \biggr] \nonumber \\
G_\theta^\theta &=& \biggl[ \frac{2r\left[rA'\ N\right]' + \left[A\ r^2 N'\right]'}{2 A\ r^2} \biggr]
\  \   = \  G_\varphi^\varphi
\end{eqnarray}
where the arguments of the functions $A(r)$ and $N(r)$ have been suppressed. For the matter field we make the following Ansatz:
\begin{eqnarray}
\Phi(x^\mu)=\phi(r) e^{i\omega t}
\end{eqnarray}
With this Ansatz, the Einstein equations of motion:
\begin{eqnarray}
\;G_t^t = 8 \pi G\ T_t^t -\Lambda~~,~~ \;G_r^r =  8 \pi G\  T_r^r -\Lambda~~,~~ \;G_\theta^\theta =  8 \pi G\ T_\theta^\theta-\Lambda~~,~~ \;G_\varphi^\varphi =  8 \pi G\ T_\varphi^\varphi-\Lambda \label{Gmn}
\end{eqnarray}
take the form:
\begin{eqnarray}
 \frac{-1}{r^2}\left[r\left(1-N\right)\right]' 
& = &-\Lambda -\frac{8\pi G}{A^2 N } \left[ \omega^2 \phi^2
+ A^2 N^2 \phi'^2 + A^2 N m^2 \phi^2 + 2 A^2 N \lambda \phi \right] 
\end{eqnarray}
\begin{eqnarray}
\frac{2 r A' N -A\left[r\left(1-N\right)\right]'}{A r^2}
& = &-\Lambda+\frac{8\pi G}{A^2 N} \left[\omega^2 \phi^2+ A^2 N^2 \phi'^2 -A^2 N m^2\phi^2 -2 A^2 N\lambda \phi \right] 
\end{eqnarray}
\begin{eqnarray}
\frac{2r\left[rA'N\right]' + \left[A r^2 N'\right]'}{2 A r^2} & = &-\Lambda+\frac{8\pi G}{A^2 N} \left[ \omega^2 \phi^2 -A^2 N^2 \phi'^2 -A^2 N m^2\phi^2 - 2 A^2 N \lambda \phi \right]  \label{eq:theta}
\end{eqnarray}
\noindent where the arguments of $A(r)$, $N(r)$ and $\phi(r)$ have again been suppressed. Also, the prime here denotes differentiation with respect to $r$. Further, the equation defining $G_\varphi^\varphi $
(cf. Eq. (\ref{Gmn})) leads to an additional equation which is identical with that of the last equation (Eq. \ref{eq:theta}).

Now we introduce the dimension-less quantities:
\begin{equation}
\alpha = \frac{8\pi G\,\lambda^2}{m^4} ~~~,~~~ \tilde{\Lambda}=\frac{\Lambda}{m^2}\label{rescaled1}
\end{equation}
\begin{equation}
 h(r)=\frac{m^2\phi(r)}{\lambda} ~~~,~~~\tilde \omega=\frac{\omega}{m}\label{rescaled2}
 \end{equation}
The equations of motion of the theory in terms of $A(r)$, $N(r)$, $h(r)$ and the above rescaled (dimension-less) parameters of the theory then read:

\alpheqn
\begin{eqnarray}
h''&=& \biggl[ \frac{A^2 N ( h+{\rm sign}(h)) -\tilde\omega^2 h}{ A^2 N^2}-\frac{2\,h'}{r}-h'\left(\frac{A'}{A}+\frac{N'}{N}\right) \biggr] ~~~\label{eq:h''}  \\
\frac{1}{r^2}\left[r\left(1-N\right)\right]'&=&~~~\tilde \Lambda+\frac{\alpha}{A^2 N}\left(\tilde\omega^2 h^2+A^2 N^2 h'^2 + A^2 N(h^2+2 h) \right)
\label{eq:G_00}\\
\mbox{\hspace{-1.5cm}}
\frac{2 r A' N -A\left[r\left(1-N\right)\right]'}{A r^2} 
&=&-\tilde \Lambda+ \frac{\alpha}{A^2 N}
\left(\tilde\omega^2 h^2+A^2 N^2 h'^2-A^2 N(h^2+2 h) \right)
\label{eq:G_11}\\
\frac{2r\left[rA'N\right]' + \left[A r^2 N'\right]'}{2 A r^2}&=& -\tilde \Lambda+ \frac{\alpha}{A^2 N}
\left(\tilde\omega^2 h^2-A^2 N^2 h'^2 - A^2 N (h^2+2 h) \right)
\label{eq:G_22}
\end{eqnarray}
\reseteqn
where 
$$ {\rm sign}(h) = \left\{ \begin{array}{ll}
\pm1 \ \ & h>0,~h<0 \\ 0 & h=0 \end{array}\right. $$

After a little bit of simplification, the above equations could be rewritten as: 
{
\alpheqn
\begin{eqnarray}
N' & = & \frac{1-N-\tilde\Lambda r^2}{r} -\frac{\alpha r}{A^2 N}
\left(\tilde \omega^2 h^2+A^2 N^2 h'^2  + A^2 N h^2+2 A^2 N h \right) \ 
\label{eq:N}\\
A' & = & 
\frac{\alpha r}{A N^2}\left(A^2 N^2 h'^2 + \tilde\omega^2 h^2\right) \ 
\label{eq:A}\\
h'' & = & 
\frac{\alpha\, r h'}{N} \left(h^2 + 2 h\right)-\frac{h'\left(1+N-\tilde\Lambda r^2\right)}{rN}
+\frac{A^2 N (h+ {\rm sign}(h)) -\tilde\omega^2 h}{A^2 N^2}\hspace{0.2cm}
\label{eq:h}
\end{eqnarray}
\reseteqn
}
To solve the above equations (i.e., the equations (\ref{eq:N}), (\ref{eq:A}) and (\ref{eq:h})) numerically, we introduce a new coordinate $~x~$ as follows:  
\begin{equation}
r= r_{\rm i} +x (r_{\rm o}-r_{\rm i})\ , \ \ \ \ 0\leq x \leq 1 \ .
\label{eq:x}
\end{equation}
which implies that $ r=r_i \ \ at \ \ x=0 $ and 
$ r=r_{\rm o}\ \ at \ \ x=1 $.
Thus the inner and outer boundaries of the shell are always at $x=0$ and $x=1$ respectively, while their radii $r_{\rm i}$ and $r_{\rm o}$ become free parameters. 

For the metric function A(r), we choose the 
boundary condition:
\begin{eqnarray}
A(r_{\rm o})=1 ~~,~~
\end{eqnarray}
where $r_{\rm o}$ is the outer radius of the shell. This fixes the time coordinate. For constructing globally regular ball like boson star solutions with finite energy, we choose
\begin{eqnarray}
N(0)=1 ~~,~~
h'(0)=0  ~~,~~ 
h(r_{\rm o})=0 ~~,~~ 
h'(r_{\rm o})=0  \label{bcstar}
\end{eqnarray}
The so-called outer solutions (i.e., the solutions in the exterior region $r>r_{\rm o}$) for the boson stars is given by 
\begin{equation}
h(r)=0~,~~~~A(r)=1~,~~~~N(r )= \biggl[ 1-\frac{2  M}{r} -\frac{\tilde \Lambda}{3} r^2\biggr] 
\end{equation}
Here M is the mass of all gravitating solutions. 
In the units that we employ, we find
 \begin{equation}
 M= \biggl(1-N(r_o) -\frac{\tilde\Lambda}{3} r_o^2\biggr)\frac{r_o}{2}
 \end{equation}
 The above equation also implies that $N(r_o)(\equiv N(r)|_{r=r_o})$ implicitly depends on $\alpha$ (through the solution of Eq.\ref{eq:N}).
We wish to emphasize here that in our present considerations, we allow $\tilde \Lambda$ to have   negative as well as positive values (including zero). 



The conserved current density of the theory is:
\begin{eqnarray}
j^\mu=-i \,\left\{ \Phi(\nabla^\mu \Phi)^*-\Phi^* (\nabla^\mu \Phi) \right\}\ 
\end{eqnarray}
Also, it is straight forward to see that $ \nabla_{\mu} j^{\mu} = 0$ and it implies a conserved charge Q for the theory as:
\begin{eqnarray}
Q &=&  -\frac{1}{4\pi}\int j^t \sqrt{-g} \,dr\,d\theta\,d\varphi  ~~ \\
j^{\,t}&=&-ig^{tt}\{\Phi\partial_t\Phi^* - \Phi^* \partial_t\Phi\}=-\frac{h^2(r) \tilde\omega}{A^2(r) N(r)}\frac{\lambda^2}{m^3}
\end{eqnarray}
where $j^{\,t}$ is the charge density of the theory. Thus we obtain: 
\begin{eqnarray}
Q =\frac{\lambda^2 \tilde\omega}{m^3} \int_{r_i} ^{r_{\rm o}} \frac{h^2}{A N} r^2 dr  ~~,~~
\end{eqnarray}

 \begin{figure}[h]
\begin{center}
\mbox{\hspace{-0.5cm}
\subfigure[][]{
\includegraphics[height=0.235\textheight, angle =0]{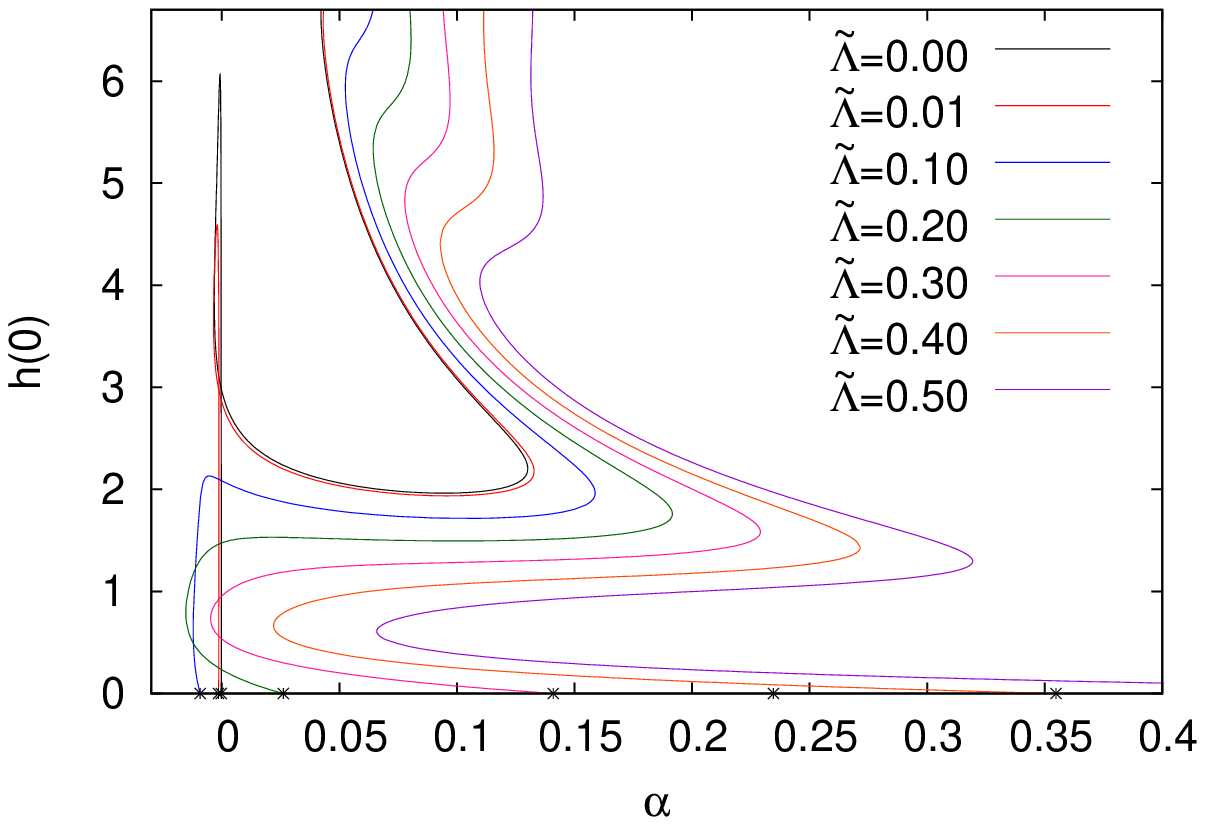}
\label{fig:hri}
}
\subfigure[][]{\hspace{-0.5cm}
\includegraphics[height=0.235\textheight, angle =0]{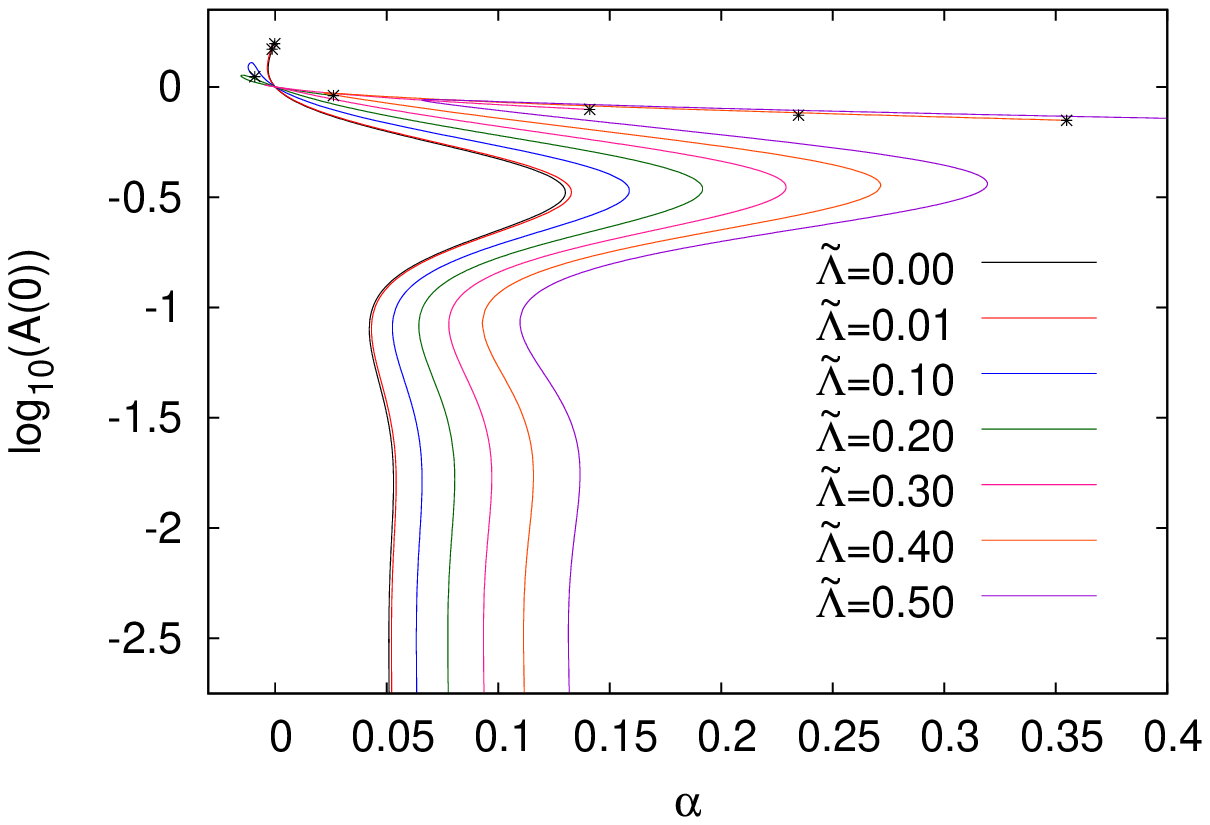}
\label{fig:lari}
}
}
\vspace{-0.5cm}
\mbox{\hspace{-0.5cm}
\subfigure[][]{
\includegraphics[height=0.235\textheight, angle =0]{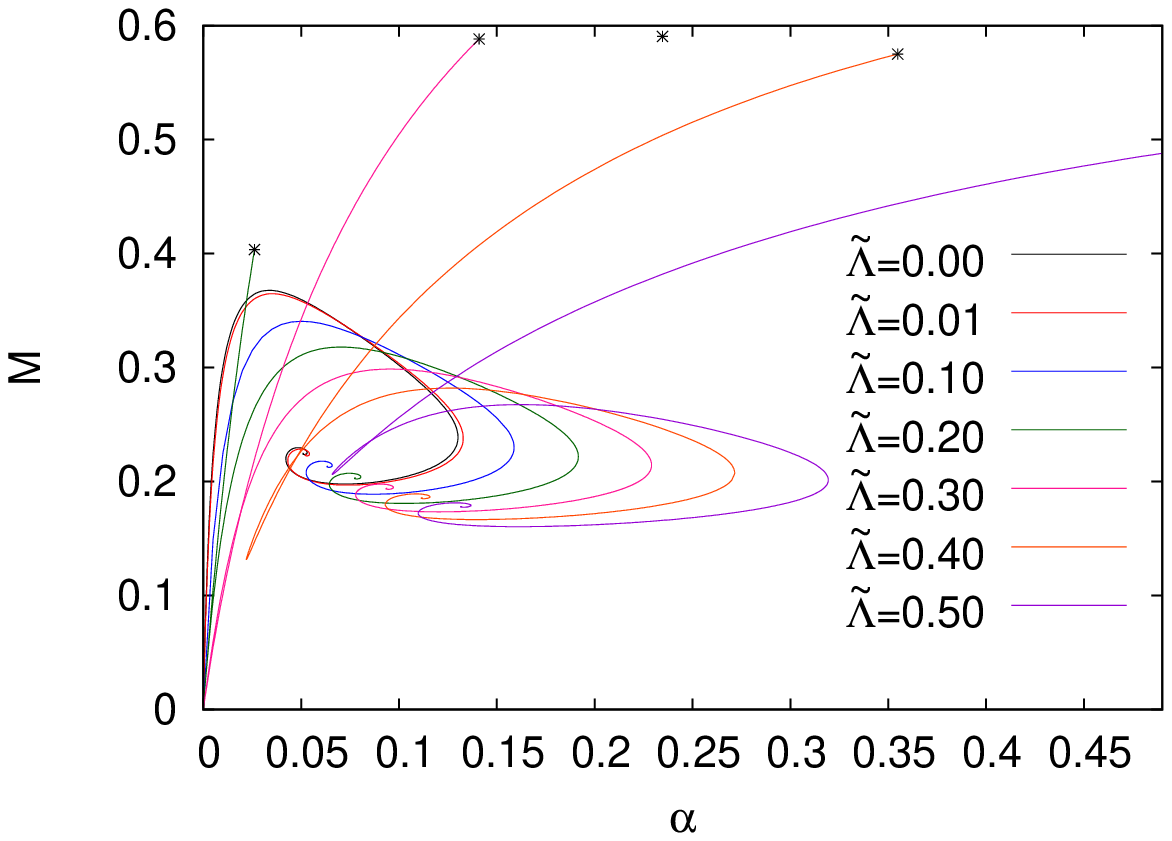}
\label{fig:Mp}
}
\subfigure[][]{\hspace{-0.5cm}
\includegraphics[height=0.235\textheight, angle =0]{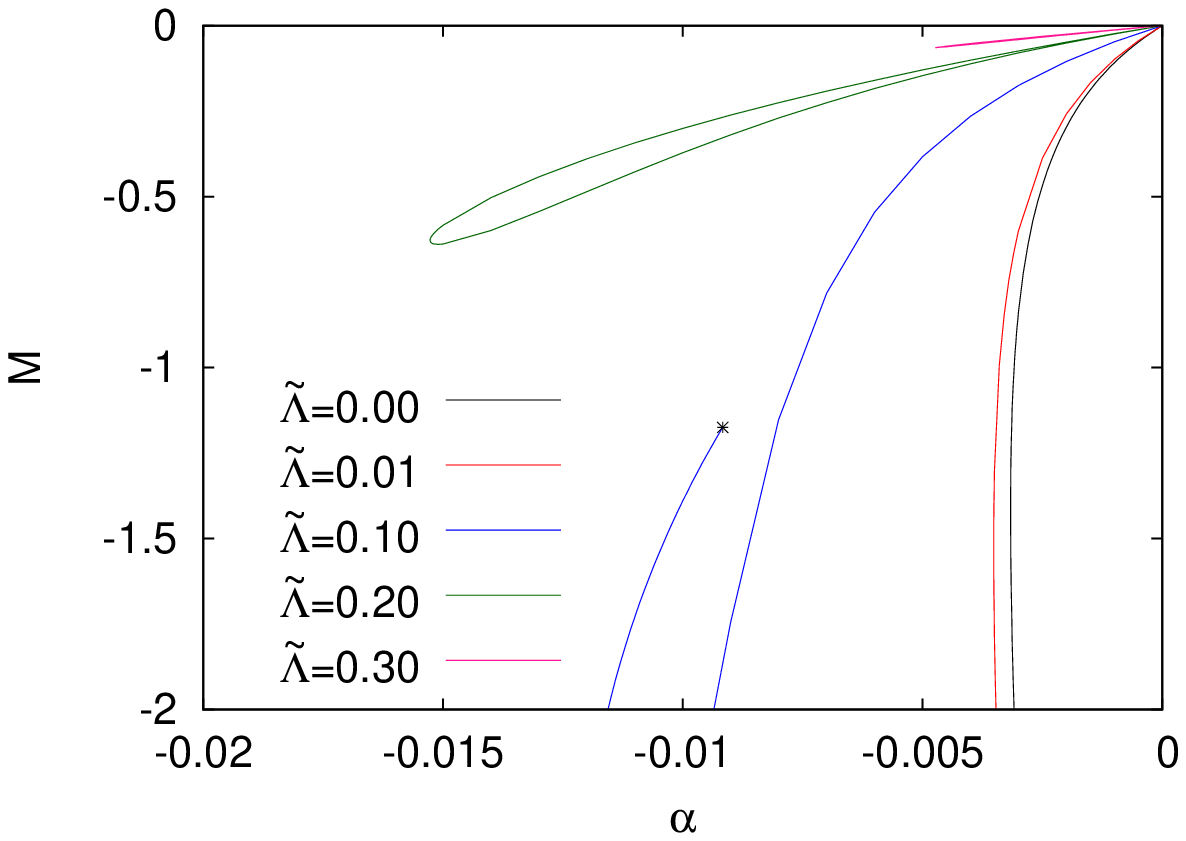}
\label{fig:Mn}
}
}
\vspace{-0.5cm}
\mbox{\hspace{-0.5cm}
\subfigure[][]{
\includegraphics[height=0.235\textheight, angle =0]{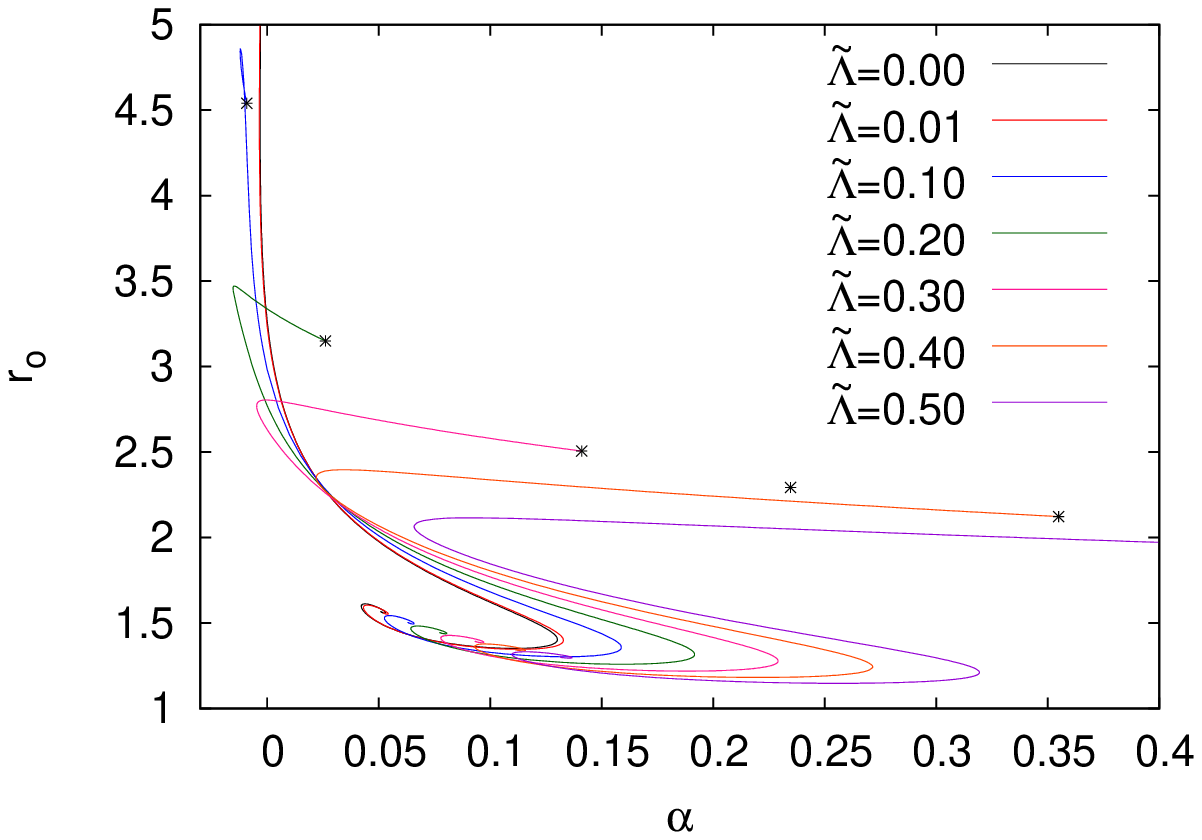}
\label{fig:ro}
}
\subfigure[][]{\hspace{-0.5cm}
\includegraphics[height=0.235\textheight, angle =0]{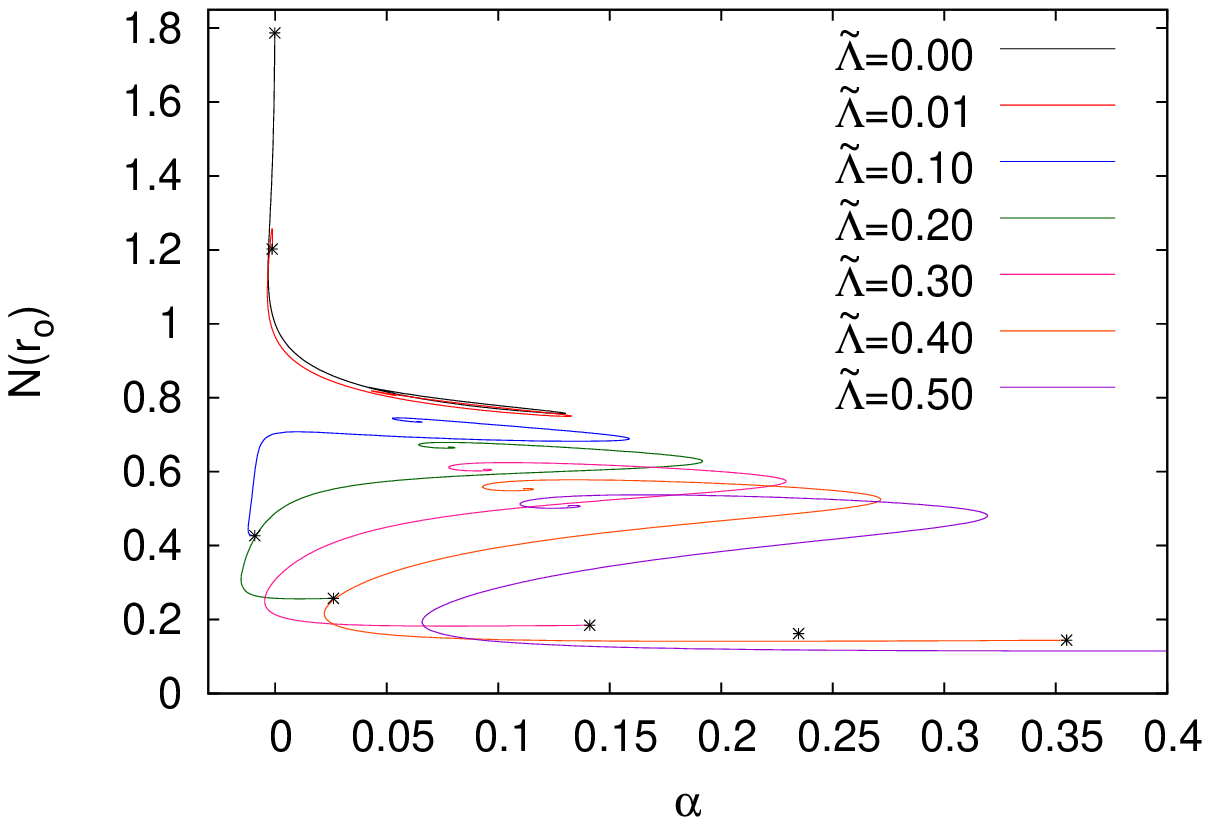}
\label{fig:nro}
}
}
\end{center}
\caption{Properties of the boson star solutions shown versus $\alpha$ :
(a) $h(0)$, the value of the scalar field function $h(r)$ at the center of star
(b) $log_{10}(A(0))$, the value of $A(r)$ at the center of star
(c) Mass parameter $M$ for positive $\alpha$
(d) Mass parameter $M$ for negative $\alpha$
(e) outer radius $r_{\rm o}$
(f) $N(r)$ at $r=r_{\rm o}$.
The asterisks mark the transition points from boson stars to boson Shells.
\label{fig:ds}
}
\end{figure}
We now briefly discuss our results for the de Sitter boson star solutions.

In figure \ref{fig:ds} we have shown the physical properties of these compact boson star solutions and the effect of the cosmological constant for several values of $\tilde \Lambda$ starting from zero. In figures \ref{fig:hri},\ref{fig:lari} we exhibit the scalar field function $h(r)$  and metric function $A(r)$ at the origin (i.e. at $r=r_{\rm i}=0$) versus coupling constant $\alpha$ ranging from negative to positive values. 

In order to see explicitly as to what happens when we look for the boson star solutions for the  negative values of $\alpha$ which, in fact, correspond to the phantom boson stars, we need to look at the terms in the action of our theory which correspond to the matter field action ($S_M$)  in terms of the rescaled scalar field (cf. Eqs. \ref{rescaled1},\ref{rescaled2}) which reads as :
\begin{eqnarray}
S_M&=&\int \mathcal L_M \sqrt{-g}\ d^4\,x\ \ \ =\int \bigl(-\partial_\mu \Phi\partial^\mu \Phi^* -V(|\Phi|)\bigr)\sqrt{-g}\ d^4 x\nonumber\\
&=&\frac{1}{16\pi G}\int \bigl(-16\pi G \partial_t (\frac{\lambda}{m^2}h(r)e^{i m \tilde\omega t})\partial^t (\frac{\lambda}{m^2}h(r)e^{-i m \tilde\omega t}) \nonumber\\ & &- 16\pi G \partial_r (\frac{\lambda}{m^2}h(r)e^{i m \tilde\omega t})\partial^r (\frac{\lambda}{m^2}h(r)e^{-i m \tilde\omega t})\nonumber\\
& &\qquad -16 \pi G \frac{\lambda^2}{m^2}(h^2+h)\bigr)\sqrt{-g}\ d^4 x\nonumber\\
&=&\frac{1}{16\pi G}\int \bigl( -2\alpha \partial_t (h(r)e^{i m \tilde \omega t}) \partial^t (h(r)e^{-i m \tilde \omega t}) - 2\alpha \partial_r (h(r)e^{i m \tilde \omega t}) \partial^r (h(r)e^{-i m \tilde \omega t})\nonumber\\
& & \qquad-2 \alpha\, m^2(h^2+h)\bigr)\sqrt{-g}\ d^4 x
\end{eqnarray}
The above equation shows explicitly the presence of $\alpha$ before the kinetic energy term of the scalar field. In view of this it becomes evident that  the negative values of $\alpha$ would give rise to negative kinetic energy for the scalar field. Such fields in the literature are generally referred to as the phantom scalar fields and the corresponding boson star solutions are called as phantom star solutions. 

When we vary $\alpha$, keeping $\tilde \Lambda$ fixed, the scalar field at the center (i.e. $h(0)$) changes continuously. At a critical value $\alpha=\alpha_{crit}$ the scalar field starts increasing  while $\alpha$ decreases. When $h(0)$ increases the coupling constant begins to exhibit damped oscillations and tends towards a limiting value $\alpha=\alpha_{lim}$. These damped oscillations in $\alpha$, lead to a spiral like behavior in other physical quantities. This spiralling behavior is a typical behavior of the boson star solutions and these spirals occur for all values of the $\tilde \Lambda$ for which the boson stars exist.

Further, in the limit $\alpha\rightarrow \alpha_{lim}$, $~ h(r)$ increases steeply at the center of the boson star while $A(r)$ decreases and tends towards zero. 

We observe that the value of the scalar field at the center depends on the cosmological constant as well as on the coupling constant of the theory. For a given finite value of $\tilde \Lambda$ the value of scalar field function $h(r)$ at the origin reaches zero at a critical value $\alpha = \alpha_{crit}$ (marked by the asterisks in figure \ref{fig:hri}). This shows the existence of the compact boson shells in the theory. For small values of $\tilde \Lambda $ (including $\tilde\Lambda$=0), the critical value $\alpha_{crit}$ is less than zero. This shows the existence of the phantom boson shells as well.

In figures \ref{fig:Mp} and \ref{fig:Mn}, we show the mass parameter $M$ versus the coupling constant $\alpha$ for its positive and negative values. The scaled mass $M$ shows the characteristic spirals of the boson star solutions. At the center of spirals, the mass parameter for the boson stars reaches its limiting value where $\alpha\rightarrow \alpha_{lim}$. The maximal value of the mass parameter is reached at $\alpha=\alpha_{crit}$

Likewise, the radius $r_{\rm o}$ of the boson star solutions as well as, the metric function $N(r)$, at the radius $r = r_{\rm o}$, versus the coupling constant $\alpha$, are shown in figures \ref{fig:ro} and \ref{fig:nro}, which show the characteristic spiralling behavior of boson star solutions. Also, the radius $r_{\rm o}$ reaches its limiting value at the center of the spirals.

The domain of existence of these boson star solutions is bounded by the lower and upper bounds on the coupling parameter (i.e. $\alpha_{min}<\alpha<\alpha_{max}$) for a given finite value of the $\tilde \Lambda$.

\begin{figure}[h]
\begin{center}
\mbox{\hspace{-0.5cm}
\subfigure[][]{
\includegraphics[height=0.235\textheight, angle =0]{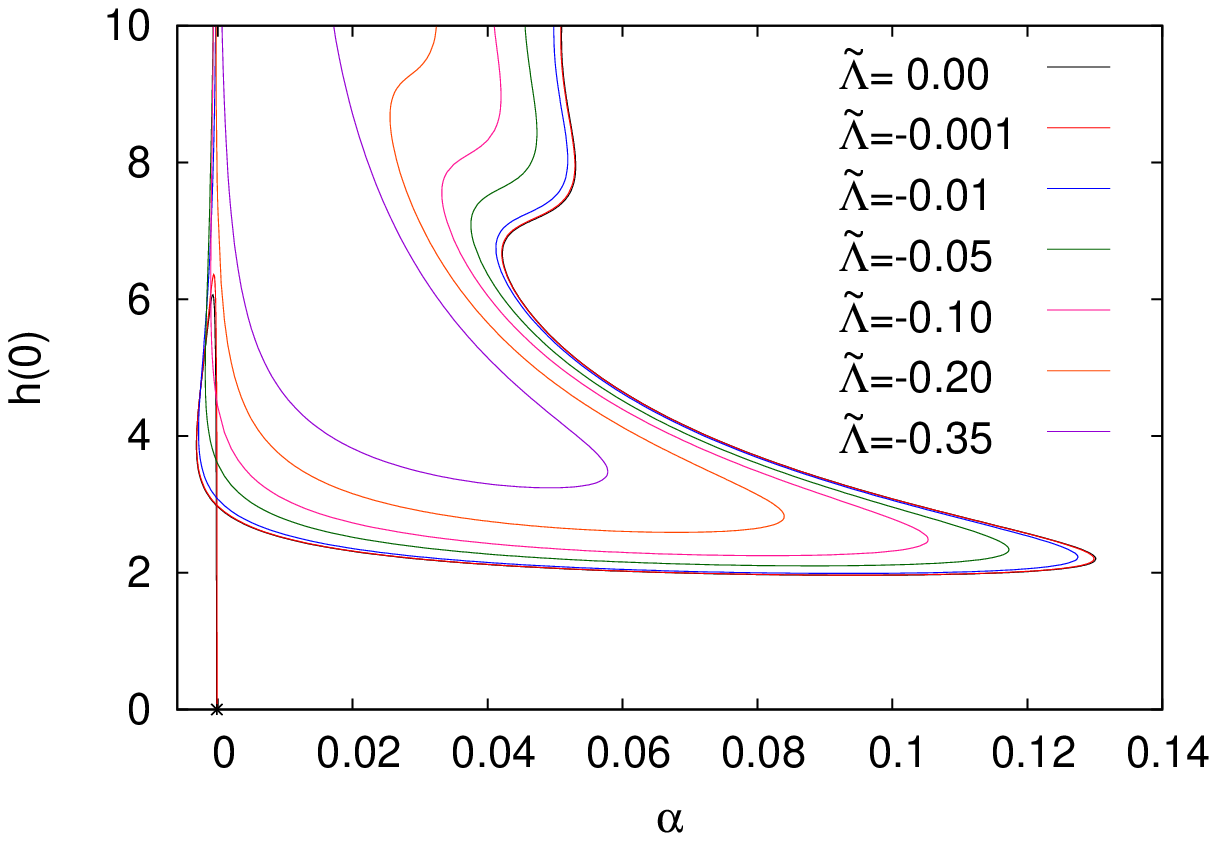}
\label{afig:hri}
}
\subfigure[][]{\hspace{-0.5cm}
\includegraphics[height=0.235\textheight, angle =0]{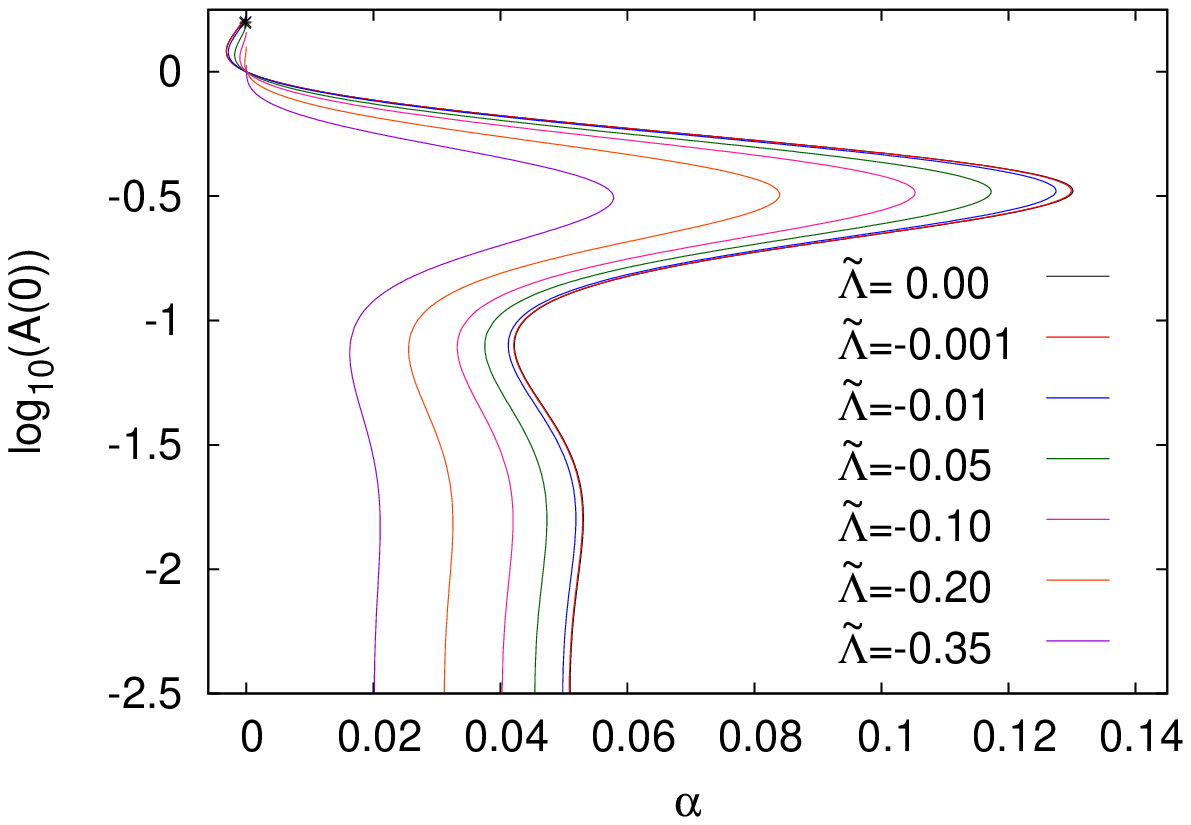}
\label{afig:lari}
}
}
\vspace{-0.5cm}
\mbox{\hspace{-0.5cm}
\subfigure[][]{
\includegraphics[height=0.235\textheight, angle =0]{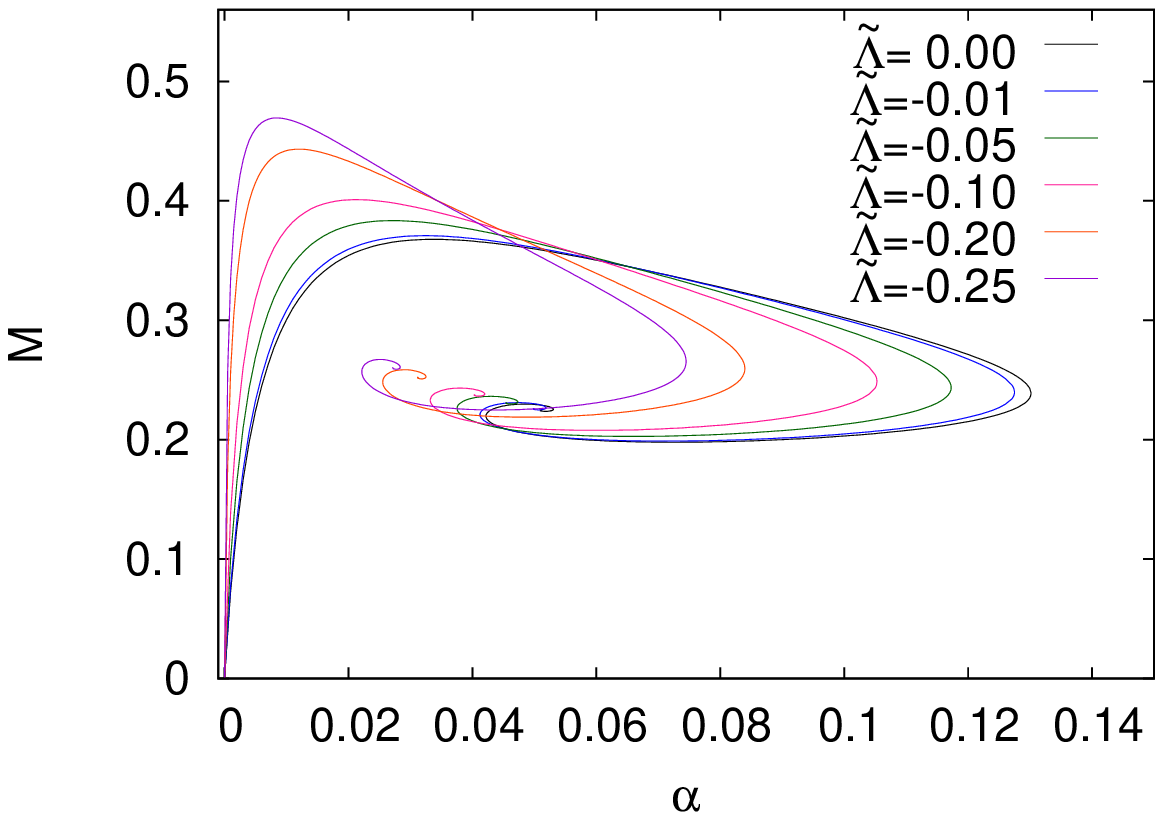}
\label{afig:Mp}
}
\subfigure[][]{\hspace{-0.5cm}
\includegraphics[height=0.235\textheight, angle =0]{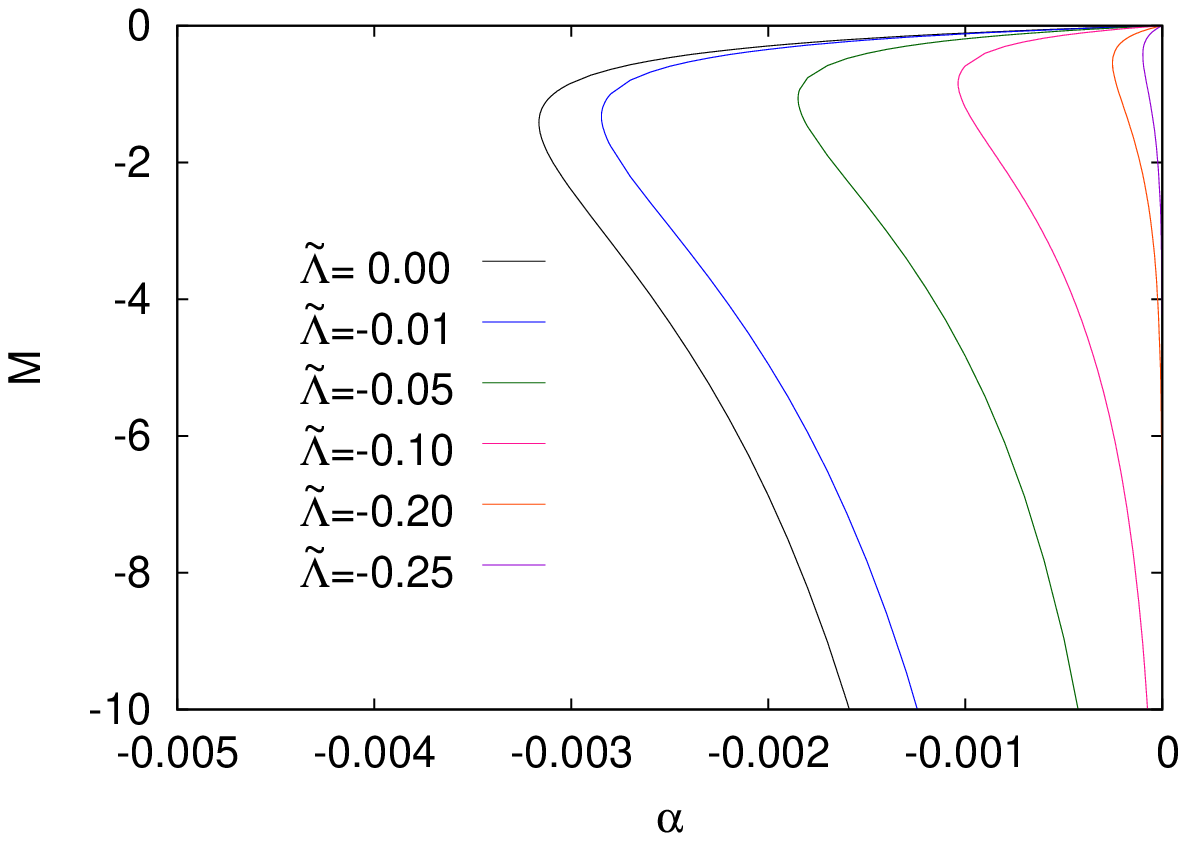}
\label{afig:Mn}
}
}
\vspace{-0.5cm}
\mbox{\hspace{-0.5cm}
\subfigure[][]{
\includegraphics[height=0.235\textheight, angle =0]{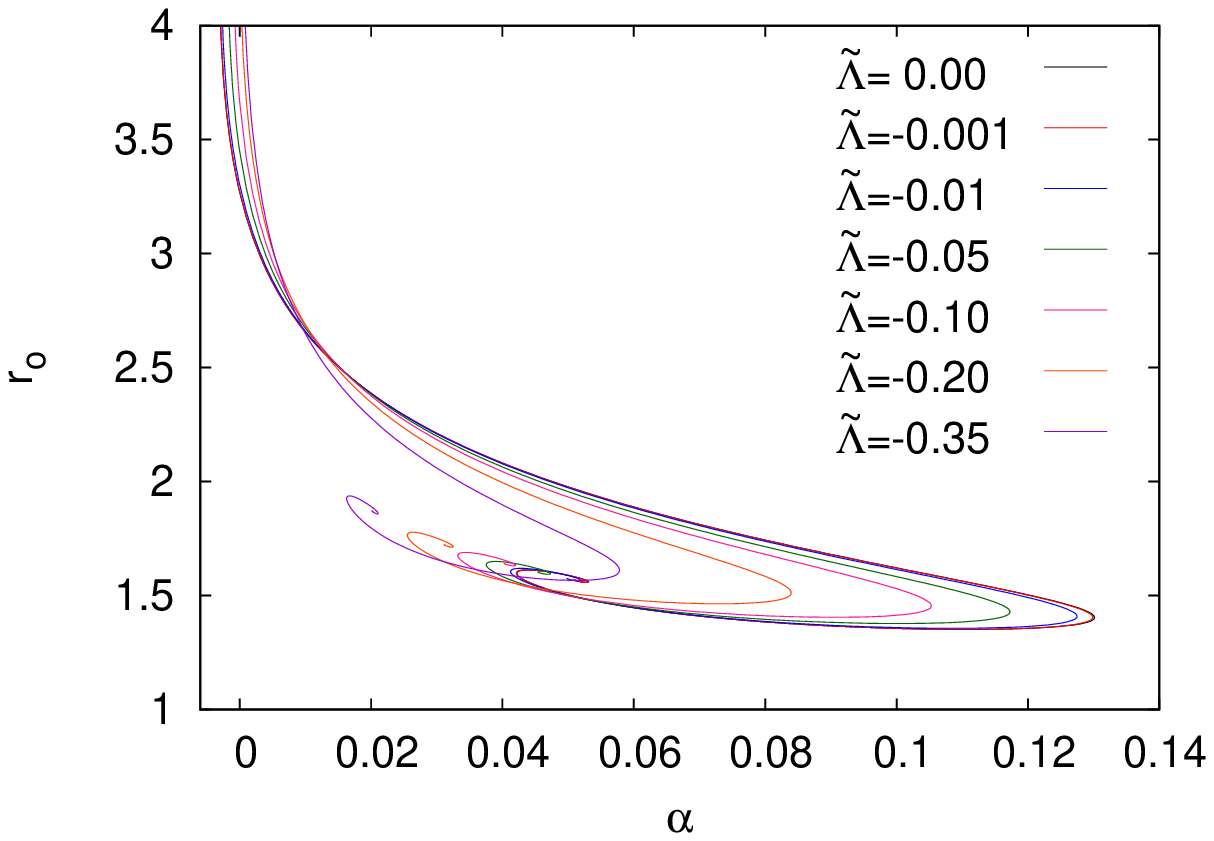}
\label{afig:ro}
}\subfigure[][]{\hspace{-0.5cm}
\includegraphics[height=0.24\textheight, angle =0]{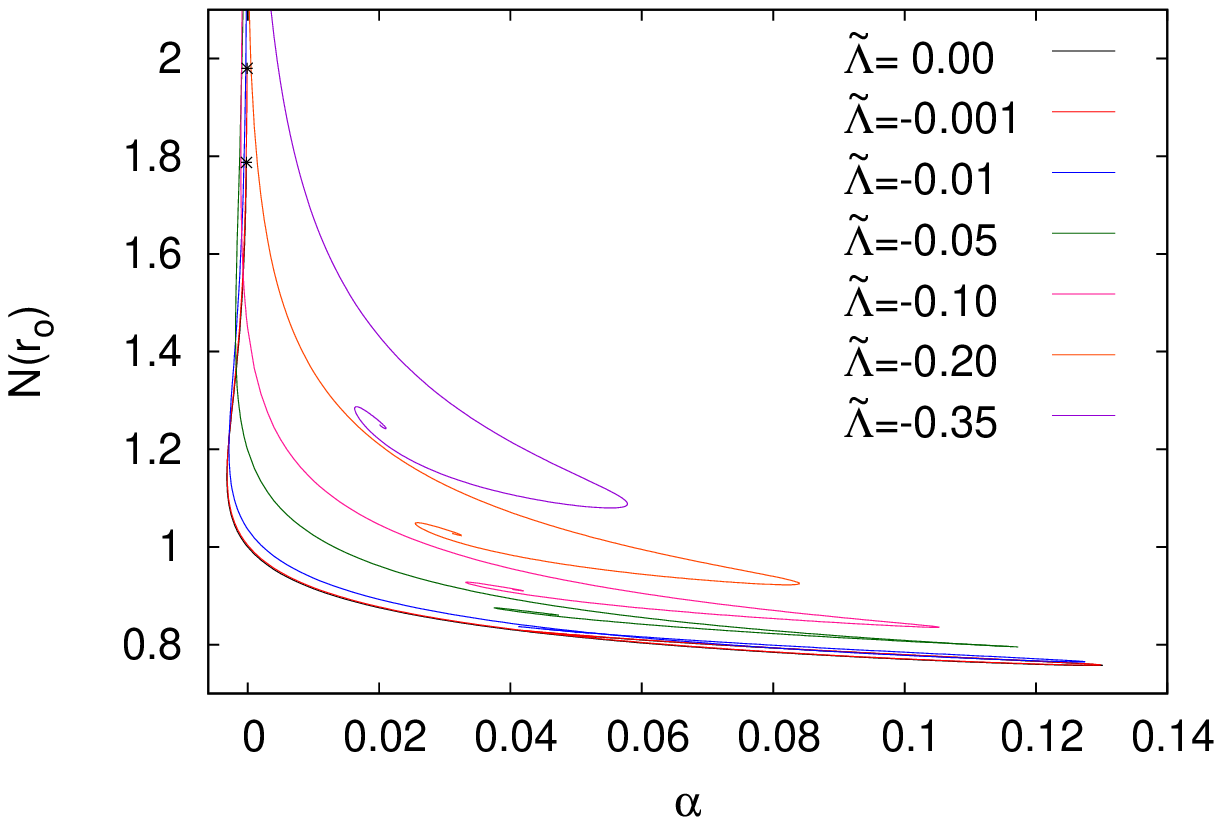}
\label{afig:nro}
}
}
\end{center}\caption{Properties of the boson star solutions shown versus $\alpha$ :
(a) $h(0)$, the value of the scalar field function $h(r)$ at the center of star
(b) $log_{10}(A(0))$, the value of $A(r)$ at the center of star
(c) Mass parameter $M$ for positive $\alpha$
(d) Mass parameter $M$ for negative $\alpha$
(e) outer radius $r_{\rm o}$
(f) $N(r)$ at $r=r_{\rm o}$.
The asterisks mark the transition points from boson stars to boson Shells.
\label{afig:ads}
}
\end{figure}

We now discuss the boson star solutions in the Anti de sitter (AdS) space. In figure \ref{afig:ads} we show some of the properties of the AdS boson  star solutions versus coupling constant $\alpha$. Here, we have extended our considerations of the solutions for the negative values of $\tilde \Lambda$ (which demonstrate that the boson star solutions also exist for the negative values of cosmological constant).

In figures \ref{afig:hri} and \ref{afig:lari}, we show the plots of the scalar field function $h(r)$  and metric function $A(r)$ for $r=r_{\rm i}=0$ versus the coupling constant $\alpha$. As we increase $\alpha$ the scalar field and the metric functions change continuously. The curve corresponding to each solution is bounded by a maximal value $\alpha=\alpha_{max}$. At $\alpha_{max}$, the value of $h(0)$ starts increasing monotonically while $\alpha$ shows damped oscillations and tends towards the limiting value $\alpha_{lim}$. In the limit $\alpha \rightarrow \alpha_{lim}$, $h(0)$ increases steeply while $A(0)$ decreases and tends towards zero.

Now as we decrease $\tilde \Lambda$, the domain of existence of boson stars also decreases. This implies that there is a limiting value for $\tilde \Lambda$ for the compact boson star solutions. Here, analogous to the de Sitter case, we find that there exist phantom boson stars associated with the negative $\alpha$ and the curve corresponding to each solution is bounded by a minimal value $\alpha=\alpha_{min}$. Interestingly we observe that scalar field at the center reaches zero for negative values of $\alpha$ for small negative values of $\tilde \Lambda$ (including the value $\tilde \Lambda=0$). This signals the presence of phantom boson shells in the theory. 

In figures \ref{afig:Mp}-\ref{afig:nro}, we show the plots of the scaled mass, radius of the boson stars and the metric function $N(r_{\rm o})$ with respect to $\alpha$. We find that the physical properties of boson stars exhibit characteristic spiralling behavior for the boson star solutions.

To summarize, we have studied the boson stars in a theory of a complex scalar field with a particular self-interaction potential (defined by equation(\ref{potential})), coupled to Einstein gravity in the presence of a cosmological constant. We have obtained the boson star solutions and phantom star solutions by analyzing the scaled equations numerically. All boson star solutions exhibit the spiral like dependence of the radius and mass on the coupling constant $\alpha$. The scalar field at the center: $h(0)$ increases while $\alpha$ exhibits the damped oscillation.

As mentioned in the foregoing, if we allow $\alpha$ to have negative values then the solutions for the negative values of $\alpha$ would actually be the solutions which correspond to the so-called  phantom scalar fields (which in turn, correspond to the negative kinetic energy for the scalar field).


Our present work could be easily extended to the case of charged boson star and boson shell solutions (analogous to the work of Refs. \cite{sanjeev:2014cqg,Hartmann:2012da}), as well as to other dimensions (analogous to the work of Ref. \cite{Hartmann:2013kna}).

\begin{acknowledgements}
The authors thank Jutta Kunz, Burkhard Kleihaus and Betti Hartmann for several educative discussions and for introducing us to this interesting subject. One of us (SK) thanks the Council for Scientific and Industrial Research, New Delhi for the award of a Junior Research Fellowship.
\end{acknowledgements}

\end{document}